\documentclass[sigconf]{acmart}

\usepackage[utf8]{inputenc}
\usepackage{booktabs}
\usepackage{multirow}
\usepackage{microtype}
\usepackage{todonotes}
\usepackage{color}

\newcommand{\nitin}[1]{\todo[inline,caption={},color=magenta!40]{{\it Nitin:~}#1}}
\newcommand{\reuben}[1]{\todo[inline,color=green!40]{Reuben: #1}}
\newcommand{\maxv}[1]{\todo[inline,color=blue!40]{Max: #1}}
\newcommand{\kim}[1]{\todo[inline,color=red!40]{Kim: #1}}
\newcommand{\nigel}[1]{\todo[inline,color=yellow!40]{Nigel: #1}}

\copyrightyear{2021}
\acmYear{2021}
\setcopyright{acmlicensed}\acmConference[CHI '21]{CHI Conference on Human Factors in Computing Systems}{May 8--13, 2021}{Yokohama, Japan}
\acmBooktitle{CHI Conference on Human Factors in Computing Systems (CHI '21), May 8--13, 2021, Yokohama, Japan}
\acmPrice{15.00}
\acmDOI{10.1145/3411764.3445677}
\acmISBN{978-1-4503-8096-6/21/05}

\begin{document}

\author{Nitin Agrawal}
    \email{nitin.agrawal@cs.ox.ac.uk}
    \affiliation{
    \institution{University of Oxford}
    \institution{Oxford, UK}
    }
\author{Reuben Binns}
    \email{reuben.binns@cs.ox.ac.uk}
    \affiliation{
    \institution{University of Oxford}
    \institution{Oxford, UK}
    }
\author{Max Van Kleek}
    \email{max.van.kleek@cs.ox.ac.uk}
    \affiliation{
    \institution{University of Oxford}
    \institution{Oxford, UK}
    }
\author{Kim Laine}
    \email{kim.laine@microsoft.com}
    \affiliation{
    \institution{Microsoft Research}
    \institution{Redmond, CA, USA}
    }
\author{Nigel Shadbolt}
    \email{nigel.shadbolt@cs.ox.ac.uk}
    \affiliation{
    \institution{University of Oxford}
    \institution{Oxford, UK}
    }

% \author{Authors Anonymous}
% \email{firstname.lastname@mail.com}
% \affiliation{%
% 	\institution{Institute for Clarity in Documentation}
% 	\streetaddress{P.O. Box 1212}
% 	\city{City}
% 	\state{State}
% 	\postcode{post-code}

\renewcommand{\shortauthors}{}



% \ccsdesc[500]{Human-centered computing~Empirical studies in HCI}
% \ccsdesc[500]{Human-centered computing~Accessibility design and evaluation methods}


% \title[shorttitlehere]{}
\title[Exploring Design and Governance Challenges in the Development of Privacy-Preserving Computation]{Exploring Design and Governance Challenges in the Development of Privacy-Preserving Computation}
% \title[shorttitlehere]{Understanding Motivations, Opportunities, Challenges, and Design Needs in Privacy-Preserving Computation}
\begin{abstract}
Homomorphic encryption, secure multi-party computation, and differential privacy are part of an emerging class of Privacy Enhancing Technologies which share a common promise: to preserve privacy whilst also obtaining the benefits of computational analysis. Due to their relative novelty, complexity, and opacity, these technologies provoke a variety of novel questions for design and governance. We interviewed researchers, developers, industry leaders, policymakers, and designers involved in their deployment to explore motivations, expectations, perceived opportunities and barriers to adoption. This provided insight into several pertinent challenges facing the adoption of these technologies, including: how they might make a nebulous concept like privacy computationally tractable; how to make them more usable by developers; and how they could be explained and made accountable to stakeholders and wider society. We conclude with implications for the development, deployment, and responsible governance of these privacy-preserving computation techniques.
\end{abstract}

\maketitle

\section{Introduction}

% somewhat inspired by elsden et al's blockchain paper intro which begins by setting out a problem in the field of HCI generally, before even talking about blockchain!

HCI research on privacy has traditionally focused on end-users: understanding their privacy attitudes and mental models, studying their privacy-related behaviours, and designing tools to help them manage data disclosure according to their preferences. While important, this paradigm of end-user privacy also has limitations. First, individuals may have their data processed in remote and opaque ways by dint of being taxpayers, credit risks, or suspected terrorists --- not `end users' as traditionally conceived. In such cases we still need to understand how privacy as a human right and public good can be reflected and governed in such systems. Second, the end-user privacy paradigm neglects the many other entities who play an important role in articulating, navigating, and embedding privacy in a range of contexts. If HCI is about reflecting human values in computer systems more broadly \cite{sellen2009reflecting}, it is equally important to study those people, whether they are developers \cite{balebako2014privacy}, designers, risk managers, or policy makers. Finally, addressing privacy as a problem of end-user interaction often yields depressing results due to the sheer complexity of personal data processing making it difficult for end-users to comprehend the choices and tools available. This is perhaps especially true when such complexity is the result of modern cryptographic techniques designed to \emph{protect} privacy \cite{whitten1999johnny}.

These three limitations are particularly salient in the context of this paper, which addresses  technologies for \emph{privacy-preserving computation}. These are a subset of Privacy Enhancing Technologies (PETs) which have emerged in recent years. These include Homomorphic encryption (HE), secure multi-party computation (SMPC), and differential privacy (DP). These foundational technologies share a common promise: to preserve privacy while also obtaining the benefits of computational analysis. HE enables computation on encrypted data, making it possible to outsource computation to another entity without them ever having access to the input data in the clear. SMPC allows multiple parties to jointly perform a computation based on multiple respective inputs without revealing those inputs to each other. DP refers to a way of measuring the extent to which the output of a computation reveals information about an individual, and a range of associated techniques for reducing it. While these technologies have been available in some form for years, and to some extent already are in deployment, recent progress in their foundational techniques and computational tractability has led some to anticipate their imminent adoption.\footnote{Recent industry analyst reports have suggested that PETs are `experiencing a renaissance'\cite{CPO2020}, and that 2020 was the `the year of PETs'\cite{del2020}}

Compared to systems and contexts typically studied in privacy-related HCI research, these privacy-preserving computation techniques may be far removed --- conceptually, operationally and experientially --- from the entities whose privacy they purport to protect. While HE, SMPC, and DP are sometimes touted in the marketing campaigns of some device makers,\footnote{See e.g. Apple \url{https://www.apple.com/privacy/docs/Differential_Privacy_Overview.pd} and Google's \cite{erlingsson2014rappor} DP initiatives} for the most part these technologies are deployed as invisible infrastructure rather than being positioned as features which end-users are expected to value, let alone understand or control themselves. In many other (actual or envisioned) deployment contexts, the data being kept private may relate to individual data subjects who are not informed or engaged with its processing; and even if they were aware, they may have no ability to distinguish between whether such processing was genuinely `privacy-enhancing' or not. Furthermore, the mathematical and computational complexity underpinning these techniques raises particular challenges to explaining them to various stakeholders; not only end users and/or data subjects, but also developers, investors, product managers, and policymakers.

These differences make privacy-preserving computation technologies a prime case study for an expanded understanding of privacy within HCI beyond traditional paradigms of user attitudes and behaviours\cite{fallman2011new}, to consider developers\cite{acar2016you}, managers, policymakers and others \cite{oudshoorn2003users}, and the roles they play in defining and operationalising goals like security and privacy. For better or worse, the development and adoption of these technologies, and the political values and consequences they reflect may ultimately have relatively little to do with `end users' as traditionally conceived. With these considerations in mind, this paper aims to explore the following:

\begin{enumerate}
    \item What challenges are associated with the adoption of privacy-preserving computation techniques for different stakeholders?
    \item What are the motivations for adopting them?
    \item Why and how should privacy-preserving computation technologies be explained, governed, and made accountable to data subjects and wider society?
\end{enumerate}

%  to examine the attitudes, mental models and behaviours of non-expert individuals in relation to computational PETs

To gain insight into these questions, we undertook a series of interviews with a variety of stakeholders involved in various ways in the development and adoption of privacy-preserving computation technologies (PPCTs). These included cryptographers and theoretical computer scientists working on foundational PPC techniques, developers of practical tools and libraries for non-expert developers, senior managers and policymakers assessing and identifying real-world use cases, practitioners building PPC products, and designers working with PPCs as a design material. Our aim was to draw out implications for HCI and design raised by this new class of technologies.
\section{Background}

We begin by briefly introducing emerging privacy-preserving computation techniques. We then situate our approach to studying them in relation to prior related work in HCI.

% defining what we mean by PETs

% Situating privacy in HCI.

\subsection{Overview of Privacy-Preserving Computation Technologies}

Privacy-preserving computation is a subset of Privacy-Enhancing Technologies (PETs). PETs are a broad category which could include everything from a sticker placed over a webcam \cite{machuletz2018webcam} to advanced cryptographic techniques \cite{menezes1996handbook}. Existing and well-established examples include encryption schemes used to secure data at rest, end-to-end encryption protecting data over the network, and anonymous routing protocols to prevent interactions between identities from being revealed. Such technologies are already widespread, embedded in products and as part of the global internet infrastructure. While they each have different underlying approaches and motivations, these technologies are primarily concerned with the protection of data, at rest and in transit. They generally assume that once data is safely transferred to a secure endpoint, it can be decrypted and computed on in the clear; that a single entity performs the computation; and that whether or not the result of the computation is `private' has a binary answer.

A more recent wave of PETs --- including homomorphic encryption, secure multi-party computation, and differential privacy ---allow these assumptions to be relaxed or even abandoned altogether. We briefly introduce them here.

\subsubsection{\textbf{Homomorphic Encryption}}
Informally, homomorphic encryption (HE) enables computation over encrypted data without ever `seeing' the input or the output. This is realized through a specific encryption and decryption scheme. In effect, a user could send their encrypted data to a service provider who could then perform the desired computation and send back the output to the user, while remaining oblivious to both the input and the output. More formally, homomorphic encryption is an encryption primitive that enables secure evaluation of an arbitrary circuit $f$ on an encryption $C(x)$ of a plaintext $x$, without decrypting $C(x)$ in the process, and without requiring any information about the private key. Such an encrypted evaluation results in an encryption $C(f(x))$~\cite{rivest1978data, gentry2009fully}, which can at a later point be decrypted by the owner of the private key, to reveal the result~$f(x)$, as if $f$ had been evaluated on the plaintext data. In principle, homomorphic encryption can be used to evaluate any circuit on encrypted data, but often a weaker functionality called leveled homomorphic encryption is used instead, which allows only circuits of a predetermined (but arbitrarily high) depth to be evaluated on encrypted data. In practice, the encryption scheme must be parameterised according to a desired depth bound of some interesting class of circuits. Homomorphic encryption, and often leveled homomorphic encryption, has found its application in problems such as secure data retrieval~\cite{yi2012single,angel2018pir,chen2018labeled,akavia2019secure}, outsourced computation~\cite{barbosa2012delegatable, kerschbaum2012outsourced} and secure machine learning as a service for sensitive data~\cite{hall2011secure,gilad2016cryptonets, sanyal2018tapas}, amongst others.

    %Gentry. HE as solving the `sysadmin problem' \footnote{https://arstechnica.com/gadgets/2020/07/ibm-completes-successful-field-trials-on-fully-homomorphic-encryption/}

\subsubsection{\textbf{Secure Multi-Party Computation}}
Secure Multi-Party Computation (SMPC) is a class of cryptographic primitives which enables secure evaluation of a function over data shared across multiple parties. It was formally introduced in 1982 as a 2 party protocol for the Millionaire's problem \cite{yao1982protocols}. Informally, SMPC primitives allow multiple parties to come together and jointly compute a function on their combined inputs while remaining oblivious to each other's inputs; the Millionaire's problem involves two parties learning which has greater wealth without revealing their respective fortunes. Formally, in an $n-$party setting, party $P_i$ possess an input $x_i$ and gets an output $y_i$ upon computation of function $f$ over the combined set of $x_is$ $(i \in \{1...n\})$. The secure computation guarantees privacy of the individual inputs  $x_is$.  Most SMPC protocols could be defined by the choice of the circuit for computing a particular function and the type of secret sharing scheme. Use cases for SMPC include secure operations over distributed sensitive data such machine learning \cite{gao2018privacy, agrawal2019quotient, mohassel2017secureml, riazi2019xonn, hussain2020tinygarble2}, genomic comparison~\cite{jha2008towards, evans2011faster} and private set operations~\cite{hazay2018oblivious, hazay2008efficient}.

\subsubsection{\textbf{Differential Privacy}}:  
Differential Privacy (DP) \cite{dwork2008differential} is a framework for sharing information based on a dataset while statistically limiting information exposure about the individuals in the dataset. More broadly, the idea of differential privacy is to deploy a mechanism where the effect of a single substitution in a dataset is very small. In effect, a query on a dataset with such a mechanism in place does not reveal anything substantial about a single individual. Differential privacy may not always be considered a privacy-enhancing technology \emph{per se}, but rather a theory for \emph{measuring} privacy in a particular way. However there are several techniques which are closely associated with differential privacy, all of which involve adding noise to results according to differentially private constraints; we therefore refer to this family of techniques loosely as differential privacy technologies. Formally, a randomized function $f$ gives $(\epsilon, \delta)$-differential privacy for all databases $D$ and $D'$, non-negative values $\epsilon$, $\delta$ and $\forall S \subseteq $ range of $f$, where $D$ and $D'$ differs by at most one record iff,
\[Pr[f(D) \in S] \leq \delta + e^{\epsilon}Pr[f(D') \in S]\]
 Here $\delta$ and $\epsilon$ are the privacy parameters. Differential privacy is one of the more widely deployed privacy-preserving computation technologies. It can be applied to querying databases\cite{johnson2018towards}, building differentially private machine learning models \cite{abadi2016deep,wei2020federated} and performing statistical analysis \cite{dwork2009differential,dwork2010differential} with privacy guarantees. More recently, the US census used DP in 2020\footnote{\url{https://www.census.gov/newsroom/blogs/research-matters/2018/08/protecting_the_confi.html}}, Apple has deployed local DP for a number of features\footnote{\url{https://www.apple.com/privacy/docs/Differential_Privacy_Overview.pdf}} and Google has been using DP for collecting data over its Chrome browser \cite{erlingsson2014rappor} in a privacy preserving manner.

\subsubsection{`Privacy-preserving computation'}\label{ppc_def}

Despite their differences, these technologies have all been classed as `tools for privacy-preserving computation'\cite{un2020ppc}, which enable `the derivation of useful results from data without giving other people access' to such data \cite{royal2019protecting}.
These privacy-preserving computation technologies are still an emerging technology. While significant theoretical progress has been made, this has yet to be translated into widespread adoption. However, numerous libraries exist for HE, SMPC and DP\footnote{For HE: Microsoft SEAL~\cite{sealcrypto}, HElib (\url{https://github.com/homenc/HElib}), PALISADE~(\url{https://palisade-crypto.org}); for SMPC: Crypten~(\url{https://crypten.ai}), emp-toolkit~\cite{emp-toolkit}, SPDZ~(\url{https://github.com/bristolcrypto/SPDZ-2}); for DP: Google's DP framework~(\url{https://github.com/google/differential-privacy}), Diffprivlib~(\url{https://github.com/IBM/differential-privacy-library}), Pysyft~\cite{ryffel2018generic}.}, the number of government-funded PPC projects is increasing \cite{un2020ppc}, and various industry and policymaking forums have publicly heralded their potential. Recent reports and working papers have catalogued actual or potential use-cases, as well as noting possible usability barriers including programming complexity, computational overhead, and parameter selection \cite{cammarota2020trustworthy,un2020ppc}. %As with any new theoretical advance, the road towards real-world uses may be long and uncertain.

\subsection{Related work on PETs in HCI}

% In this section we draw together a range of background literature from HCI, usable privacy and security and situate our approach among it.

Much research on privacy in HCI is concerned with how end-users value, negotiate, and manage privacy in the context of their interactions with computers. Work in this vein involves: understanding the attitudes \cite{ur-2012-smart,king2013come,shklovski2014leakiness,van2017better,kumaraguru-2005-privacy}, expectations \cite{lin2012expectation, balebako2013little} and mental models \cite{kang2015my} of end-users regarding how their data is collected and used; studying privacy-related behaviours such as willingness to share data \cite{leon2013matters,acquisti-2015-privacy} and use of protective measures \cite{baruh-2017-online}; and evaluating and designing tools for privacy management such as permission settings \cite{leon2012johnny,lin2014modeling}, privacy notices \cite{schaub2015design, balebako2015impact}, and privacy assistants \cite{liu2016follow}. Related research in usable privacy and security addresses the usability of various end-user PETs tools. These include privacy and security aspects of ubiquitous tools e.g. web browsers \cite{herzberg2009johnny,cranor2002use}, as well as more advanced specialist tools, such as end-to-end encryption \cite{whitten1999johnny} and anonymous communication and routing tools \cite{clark2007usability}. Such work is highly relevant to contexts in which end users directly interact with systems in ways that may affect their privacy, and where there are opportunities to (re)design tools and interfaces to give them more control. Such work is premised on the ideal of individual users being able to understand at least some aspects of how their data is processed, and having the potential to exert some meaningful choices over it.

In some cases, privacy-preserving computation technologies might be usefully studied from this end-user perspective. Bullek et al. \cite{bullek2017towards} studied people's comprehension of the randomized response method for local differential privacy \cite{warner1965randomized}. Participants were asked a series of questions, the answers to which were perturbed with noise to provide privacy. In response to a final question about a particularly sensitive topic, they were able to choose how much perturbation to add (i.e. the value of $\epsilon$). While most participants selected the lowest (most privacy-preserving) value for $\epsilon$, surprisingly, 20\% chose the highest (least privacy-preserving) value for $\epsilon$. Some participants explained this was because adding more noise felt like lying. Xiong et al. \cite{xiong2020towards} also studied participants' willingness to share data with a hypothetical differentially private system. They examined the effect of different descriptions of differential privacy (including real descriptions provided by technology companies and the U.S. Census bureau) on willingness to share, and their findings suggest that certain descriptions (in particular, implication descriptions) are more understandable and increase willingness to share data as a result. Finally, Qin et al. explored usability and understanding in the context of privacy-preserving data aggregation initiatives based on MPC, finding that using various analogies to explain the process of additive secret sharing increased participants' confidence in the scheme \cite{qin2019usability}.

However, in many contexts, the `data subjects' are not co-extensive with the `users'. In the case of the PPCTs mentioned above, there may be several primary users (which may include developers and others) and many wider ‘stakeholders’ (e.g. commercial and government partners, the wider public). Rather than studying end users who are also data subjects, then, we might instead follow previous HCI research on privacy and security which focuses instead on other actors, such as developers (e.g. \cite{balebako2014privacy,acar2016you, gorski2020listen,assal2019think}). Balebako et al. note that while users may be concerned about privacy, they are generally not `empowered to protect themselves'; by contrast `the decisions made by app developers have great impact' \cite{balebako2014privacy}. Studying developers, designers, and others can reveal both practical and organisational challenges hindering the deployment of privacy and security technologies \cite{acar2016you,furnell2009integrated}, highlight discrepancies between privacy research and privacy engineering \cite{gurses2017privacy,kostova2020privacy}, as well as elucidate the moral dimensions of design. While some studies of software developers suggest that they may `not have sufficient knowledge and understanding of the concept of informational privacy'~\cite{hadar2018privacy}, others do explicitly engage with the ethical and political ramifications of their work; e.g. Rogaway~\cite{rogaway2015moral} who acknowledges how the field of cryptographic privacy technologies has `an intrinsically \textit{moral} dimension'.

This kind of reflexivity on the part of developers and designers is something acknowledged and addressed in approaches like Value Sensitive Design (VSD), which aim to `illuminate the ethical and moral responsibility on the part of the designer rather than the user' \cite{fallman2011new}. To understand how particular technologies are imagined as solutions to problems \cite{jasanoff2009containing}, we may need to study a wide variety of actors involved in their development, not only engineers but also those involved in the business of marketing them \cite{oudshoorn2003users}. By encompassing the full breadth of different actors involved in creating and deploying these systems, we are also able to grapple in different ways with the trade-offs and tensions inherent in the field of privacy-preserving computation, and ask questions like `“Who is making the design decision?”, “Who is paying for it?”, “What is this saying about the user?”' \cite{harrison2007three}.

% cite design tensions \cite{tatar2007design} ?

% privacy mindset of software developers \cite{hadar2018privacy} `classified and analyzed the cognitive, organizational and behavioral factors that play a role in developers’ privacy decision makin' `organizational privacy climate is a powerful means for organizations to guide
% developers toward particular practices of privacy; and that software architectural patterns frame privacy solutions that are used throughout the development process' `many developers do not have sufficient knowledge and understanding of the concept of informational privacy'

% % to understand the design and governance challenges of computational PETs.

% In moving beyond end-users and usability, to 

% Value sensitive design - values have `moral epistemic standing' \cite{friedman2003human} `VSD comes to illuminate the ethical and moral responsibility on the part of the designer rather than the user' \cite{fallman2011new}
% humanistic \cite{bardzell2015humanistic}. All paradigms are valuable, and HCI can combine multiple \cite{duarte2016revisiting}.

Finally, there is also work which critically addresses Privacy-Enhancing Technologies from a philosophical and conceptual perspective. For instance, Tavani and Moor~\cite{tavani2001privacy} assess how earlier PETs such as PGP and anonymity tools may address privacy as individual control, but do not provide `external' control beyond the user, which they argue is necessary to protect privacy in the round. Gurses and Berendt point to the limitations of PETs that stem from understanding privacy solely in terms of confidentiality~\cite{gurses2010pets}. Stalder points to the ways that PETs designed for individual use may occlude broader social meanings of privacy \cite{stalder2002failure}, while Phillips notes how PETs designed to assist businesses with automating compliance with privacy laws reinforce a restricted notion of privacy as unwanted intrusion \cite{phillips2004privacy}.

\section{Research Approach}
Given that the technologies being addressed here are still emerging, and the broad and exploratory nature of our research questions, we chose to undertake in-depth semi-structured interviews with a select range of experts from a range of backgrounds and roles \cite{bogner2009interviewing}. All had direct experience of working on projects relating to privacy-preserving computation, and occupied different strategic positions in the developing ecosystem. They included: researchers working across HE, SMPC and DP research; industry practitioners and designers with experience delivering practical applications of these technologies; as well policy experts with experience in PETs. We deliberately selected some experts whose careers and roles bridged between the domains of research, industry, or policy, some having moved from one to the other over the course of a career, while others maintaining feet in multiple domains simultaneously. These participants can be seen as `boundary workers', working between the boundaries of science and policy to facilitate the co-production of knowledge and innovation \cite{hoppe2009scientific} and `knowledge brokers' who facilitate connections between scientific and other audiences \cite{meyer2010rise}. Including a variety of different roles also reflects the nature of these technologies as `use-inspired basic research'\cite{oulasvirta2016hci} operating between `basic' and `applied' research paradigms \cite{stokes2011pasteur}. This enabled us to not only understand how the knowledge surrounding these technologies is made in specific places (e.g. research labs, technology companies, government) but also `how transactions occur between places'\cite{shapin1998placing}.

% the role that knowledge brokers play in articulating and translating PETs. challenge typical forms of HCI research. articulation of system requirements, design goals, etc., happen as much in the process of knowledge brokering as in the minds of researchers.

% HCI as a problem-solving discipline. \cite{oulasvirta2016hci}

% PETs as an example of `use-inspired basic research' -  where a problem in the world
% shapes the hypotheses for study in the laboratory. vannevar bush basic and applied. \cite{stokes2011pasteur}

% “We need to understand not only how knowledge is made in
% specific places but also how transactions occur between places,” Shapin (1998, pp. 6-7).

% drawing from a very small population of those who have worked in this space.

% all interviewees were to some extent professionally invested in PETs, therefore we lacked strongly critical voices. nevertheless most were also able to be critical.

Because these technologies are still emerging, we inevitably could only draw from a small class of professionals, whose roles in the production of these technologies are to some extent ill-defined. As is typical with expert interviews, there was no comprehensive list of relevant experts to sample from; we therefore built a `sample frame' based on publicly available materials from a wide variety of sources including research papers, industry and policymaking fora, and press \cite{goldstein2002getting}, to identify potentially relevant experts, and also used snowball sampling.  % centralised lists of companies deploying. to get a broad range of perspectives and roles - i.e. not only academic cryptographers - we tried to follow different leads based on research of publicly available documentation, and snowball sampling. ppts willing to disclose colleagues in other firms or public policy roles. direct enquiries with organisations likely to or already known to be engaging in PETs. Searches for related terms (HE, SMPC, DP) in news press in e.g. LexisNexis.
As a result of the variety of roles and experiences of our 9 experts (see Table \ref{tab:participants}), we used a semi-structured interview format slightly tailored to four different roles (research, industry, policy, design). We invited participants to discuss their experiences, motivations, perceived opportunities and barriers, relating to this space. Open ended questions allowed us to have relatively free reign to explore these issues \cite{page2005policy}. The interviews were conducted over video chat during Spring and Summer 2020. The 9 interviews varied in length between 35-75 minutes, with the average taking 55 minutes, producing 8.3 hours of audio recordings in total, which were transcribed. All parts of the study were approved by our institution's ethics review committee.
We used thematic analysis \cite{braun2006using} to identify key themes and ideas discussed by experts. Two of the authors independently developed a set of codes based on close reading of disjoint subsets of the interview transcripts, using an open coding process. The two authors then discussed and consolidated their codes to derive a common set, which was then applied by both authors to all the interview transcripts, and memo notes were taken to record observations about the codes and their relation to one another \cite{lempert2007asking}. A final round of discussion based on this data resulted in a set of themes and sub-themes, presented in the following section.

% Breakdown of academia, industry, policy, design
% Breakdown of locations: UK, US, Switzerland, India
% Breakdown of gender: [only 2 female?]

\begin{table}
\small
\centering
\begin{tabular}{p{0.2\linewidth} p{0.7\linewidth}}
\hline
Participants & Description (\& years experience in PETs)                                                                                                                                                                   \\ \hline
P1[R]       & Cryptography Researcher and an  industrial PETs library developer (20-30y experience)% of experience in academia + industry)        
\\
P2[R]      & Cryptography Researcher at company specialising in privacy preserving computation (5-10y experience)% in this academia + industry)                       
\\
P3[R]       & Cryptography Researcher and an industrial PETs library developer (5-10y experience)% in academia + industry)                                                            
\\
P4[P]       & Law and Policy professional working on industrial adoption of privacy-preserving ML (4-6y experience) %in academia + industry + government / non profit) 
\\ 
P5[P]       & Senior government adviser on technology, with strong interest in PETs and their applications (5-10y experience)% in this domain)                          
\\
P6[I]        & Security and Privacy Researcher working in executive role at large tech company (2-5y experience) % in this domain)                                       
\\
P7[I]         & Data Scientist working on privacy at a `big four' accounting firm (5-10y experience)                                                                         \\
P8[I]         & Researcher at consultancy specialising in privacy preserving computation as a service (2-5y experience)
\\
P9[D]      & Designer at a tech and design agency specialising in ethical use of data \& AI (2-5y experience)                                            \\ \hline
\end{tabular}
\caption{A summary table of the total participant sample from Research (R), Policy (P), Industry (I) and Design (D).}
\label{tab:participants}
\end{table}
\begin{comment}
\begin{table}[]
\begin{tabular}{@{}ll@{}}
\toprule
Participants & Description \\ \midrule
P1 (R)       &             \\
P2 (R)       &             \\
P3 (R)       &             \\
P4 (P)       &             \\
P5 (P)       &             \\
P6 (I)       &             \\
P7 (I)       &             \\
P8 (I)       &             \\
P9 (D)       &             \\ \bottomrule
\end{tabular}
\caption{A summary table of the total participant sample from Research (R), Policy (P), Industry (I) and Design (D).}
\label{tab:participants}
\end{table}
\end{comment}

\section{Findings}

We divide the findings from our interviews into two main areas. The first addresses the technical challenges and opportunities around the adoption of privacy-preserving computation techniques - primarily concerning their transition from theoretical research into practical application. The second concerns the motivations and goals for deploying these techniques to address commercial and societal goals, and how the institutions that deploy them might explain and be held accountable for their use.

% We organise our findings along three major themes: theory, practice and infrastructure; economic and regulatory conditions; and understanding, explanation and accountability.

% \subsection{Theory, Practice and Infrastructure}
% The first theme emerging from the interviews related to the challenges inherent in moving from theoretical and scientific work on PETs, into the realm of engineering and deploying them in practice.

% barriers between theory and practice
% motivations
% explanation and accountability

\subsection{Technical challenges and opportunities}

\subsubsection{From Theory to Practice}

Many participants brought up how advances in the theoretical grounding on which privacy-preserving computation techniques are based may not translate straightforwardly into specific real-world applications. This was acknowledged by both research scientists working on those foundations, and practitioners attempting to deploy the technologies in particular contexts. Many participants expressed confidence that those theoretical advances would translate into practice in time. P3, a researcher, argued that while it is \emph{`early stages, from the business development perspective of homomorphic encryption'}, he was nevertheless \emph{`confident that [the] technology is useful, practical'} (P3[R]).

It was generally accepted that much of the research had only reached the stage of proofs-of-concept rather than deployments; while \emph{`doable in principle'}, \emph{`there is a lot of work to do before that bigger picture potential can be realized'} (P1[R]). Similarly, P9[D] explained that as a designer she was \emph{`trying to make them more widely understood within the design and tech community ... There's lots of research in academia at the moment, but not very many examples of them being used in practice'}.

The policy experts we interviewed were optimistic that the technology was already nearly ready for practical deployment. For P4[P] \emph{`the technology has scaled to the point that ... it's definitely commercially deployable'}. For P5[P], while practical deployment would require a series of \emph{`reasonable engineering and architectural compromises'}, he was still optimistic that \emph{`existing approaches to homomorphic encryption are tractable'}.

While both research scientists and policy experts were optimistic about the big picture, those trying to bridge theory and practice on the ground expressed frustration that much of the research was not directly relevant. In some cases, this was because the scientific work made simplifying assumptions that were rarely satisfied in real use cases. In the context of trying to apply differential privacy techniques to a project involving time series data, P7[I] admitted that he \emph{`really struggled, you know, seeing the value in all those techniques that academia likes to talk about ... how am I going to use that with time series?'}.

Similarly, several participants pointed to the variety of messy underlying data and software issues that exist on the ground that hamper deployment. P7[I] explained that the initial challenges are around \emph{`how do we list data assets, and manage access, at an enterprise scale?'}; while P8[I] spoke of clients with various custom systems and data formats, so \emph{`while we're solving the security aspect of the communication between counterparties ... we still haven't fixed this engineering problem - it's not going away'}.

P2[R], a research scientist who had worked on both foundational theory and engineering, explained how applying techniques in practice involved moving carefully between theory and engineering:

\begin{quote}
    `Engineers do need to learn a lot to deploy these kind of technologies and what makes the whole thing complicated is that they need to acquire a kind of knowledge that is not something that the university professor knows ... a lot of low level optimizations make a huge difference, and yeah, from theory to practice they need to somehow be invented.'
\end{quote}

In P2's opinion, such work would not come from a \emph{`linear transmission of knowledge'}, but rather through continuous iteration and \emph{`course correction'} between theoreticians and engineers.

\subsubsection{Interdisciplinarity and translation between roles}

As well as theoreticians and engineers to bridge the gap between theory and practice, several participants discussed the need for people with different skills, backgrounds and motivations to work together. This included not only the combinations of different expertise involved in foundational privacy-preserving computation research such as math, statistics, and cryptography, but also specialists in specific application domains. As P2[R] noted, successful deployment depends on a \emph{`component of multidisciplinarity'}:

\begin{quote}
    `In my experience this is very hard to get the right people and crucially, with the right incentives in the same room ... You not only need data scientists, but also security engineers, mathematicians and then experts in the application domain.'
\end{quote}

% alternative:
% `For example, the algorithmic optimizations at the application layer that still requires deep expertise. So working with some crypto experts ... make it much easier ... for the machine learning experts to use this technology'

The differing motivations and cultures of these different communities was seen by some as as a problem as it leads to certain important problems being neglected. Commenting on the misalignment between incentives of academic researchers and industry, P3(R), who had worked in both sectors, lamented how research is \emph{`driven by the need to get published; it favors more ... performance breakthroughs, functional breakthroughs'}, meanwhile, topics like usability are \emph{`not so interesting for the basic core research community'}.

The need for an even broader range of disciplinary expertise and professional skills was articulated by P4[P], who described her role as \emph{`to bridge the lexical gap between technologists, lawyers, and policymakers to defragment the current initiatives in PETs'}. Drawing from previous experience working on AI in government, where \emph{`insulated development'} led by technologists failed to account for the \emph{`constitutional implications'} of these technologies, she warned that \emph{`the same could happen for PETs without this sort of ... interdisciplinary discourse'}.

% Furthemore:
% How people in different roles define PETs
% Translating expertise between roles

\subsubsection{Usability for Developers}

A common theme among both researchers and industry practitioners was the complexity of applying privacy-preserving PETs from a software engineering perspective. They discussed a set of inherent challenges facing developers around flexibility, performance, and specifying appropriate parameters.

Several participants described how PPCTs, in particular homomorphic encryption, can be very \emph{`brittle'} (P1[R]): small changes in parameters can result in drastic reductions in performance, security, or privacy guarantees. Such sensitivity can be hard for developers to anticipate and manage, especially as there are many different parameters to tune. This was contrasted against other PETs, like public key cryptoschemes, where there is one main parameter --- key size--- which has a fairly predictable relationship with security guarantees and computational overheads:

\begin{quote}
    `I mean RSA, you have the bit length and that's pretty much it. These things [homomorphic encryption applications] you have a ton of decisions to make when it comes to how to instantiate it, and they have implications for both speed and for the actual function that you will need to compute.' (P1[R])
\end{quote}

This results in problems for developers not just in the initial implementation of a privacy-preserving technique, but also as they inevitably need to update a system to \emph{`evolve when you need to change various details ... performance and tractability can be so highly dependent on small details'} (P1[R]).

Participants articulated this as a trade off between approaches to building PPCTs that either work out-of-the-box but have poor and unpredictable performance, \emph{or} that have reasonable performance, but require fine-tuning by engineers. For instance, while the invention of fully homomorphic encryption enables both addition and multiplication and therefore arbitrary computation, specific applications still need to be converted into those arithmetic operations and may incur great computational costs depending on how that is implemented. While it may be possible to `come up with a system with adequate performance for your application', this often requires having an application which is \emph{`fully specified and well defined in mind, and you have a team of experts working for you'} (P1[R]).

Several participants spoke about the development of privacy-preserving computation software libraries for developers (see \ref{ppc_def}), often contrasting two approaches which reflect the trade-off articulated above: on the one hand, libraries which create an abstraction layer which obscures the underlying complexity; and on the other, libraries which expose all of that complexity so that the developers still need to create bespoke solutions for their application context. P1[R] noted that many developers expect a library to provide \emph{`abstractions that are convenient'}; otherwise \emph{`It's like telling people: OK, I'll give you transistors and you'll build from them ... people don't think this way and for good reason'}. P2[R] made a comparison to machine learning frameworks (e.g. Tensorflow and Scipy):

\begin{quote}
They have a very nice abstraction layer that allows them to say "OK, here's my function over the reals: optimize it, under the hood"... We will have worried about implementing matrix multiplication super quickly over floating point numbers so that the data scientists can assume [it's] like doing math on their on their notebook, right? This level of abstraction doesn't exist yet [for privacy-preserving computation]
\end{quote}

However, P2[R] cautioned against such an approach for privacy-preserving computation libraries, because it would preclude \emph{`a lot of optimizations that come from understanding the underlying protocol'}. P1 echoed this, stating that: 
\begin{quote}
`The only way that we know now of making the computation go reasonably fast is to use a lot of tricks and the developer needs to know about those tricks'
\end{quote}

As a result, P2[R] felt that \emph{`general purpose tools'} would inevitably fail to meet developer's performance expectations and thus give them a mistaken impression about the true potential of PETs.

While many participants were in favor of some form of standardisation via libraries, P8[I] explained that the prospects for a standard platform depends on where the technologies are being deployed to. In the context of SMPC, because smartphone operating system providers \emph{`control the platform, they can decide ... this is how it's going to work'}; whereas P8[I]'s work involved deploying SMPC into a wide range of different clients' environments where \emph{`we can't really dictate to them how they store their data'}; as a result the possibility for standardisation was small.
% On this approach, tools for privacy-preserving computation techniques could never fully abstract away the details of implementation.

% \subsubsection{Design challenges}
% In addition, P9

\subsection{Motivating, explaining and governing privacy-preserving computation}

Our participants also raised various insights relating to the motivations for adopting PPCTs, and challenges relating to explanation and accountability.

\subsubsection{Motivations}
Unsurprisingly, `privacy' was often cited as the motivation for developing and deploying privacy-preserving computation technologies. However, some subtly different articulations and understandings of privacy emerged from our interviews, as well as some other motivations which went beyond privacy altogether.

Some very directly motivated the adoption of privacy-preserving computation by reference to the interests of individuals in privacy and the protection of their personal data: \emph{`it's individual privacy - it's human rights'} (P6[I]). In comparison to the push for similar technologies in other markets, privacy-preserving computation was more a response to individual privacy: 

\begin{quote}
`People do understand when their privacy is violated. So ... the push for these technologies is very different to the push the semiconductor industries have had ... So I give a lot of time on the examples that are user-centric' (P6[I]).    
\end{quote}

% `there is one thing that is very different from e.g. HE and RSA. people do understand when their privacy is violated. it's their need to keep their privacy. so therefore the push for these technologies is very different to the push of the semiconductor industries have had before to bring security technologies based on cryptography into the market. so i give a lot of time on the examples that are user centric'

However, appeals to individual privacy were often mediated via other pressures. First, organisations deploying PETs may not have a direct relationship with those individuals, but are instead concerned with third-parties affected through business-to-business relations:

\begin{quote}
    `You have customers: these may be business to business customers, but that also extends to customers of customers and therefore it boils down to individuals' (P6[I])
\end{quote}

Second, some cited the existence of privacy and data protection regulation as an incentive to provide and deploy PETs: \emph{`because of GDPR} [the E.U. General Data Protection Regulation]\emph{... all the regulatory environment is ... very favorable for providers'} (P7[I]). This regulatory pressure meant that investment in PPC could be accounted for in terms of corporate risk management: \emph{`to have compliance at least formally speaking with GDPR ... it's really protecting assets of a company'} (P6[I]).\footnote{A sentiment echoed in \cite{cammarota2020trustworthy}} % `there is a whole spectrum of things and on the other end there are regulators'

Third, P9[D] argued that rather than just enabling existing data processing to be done in a more privacy-preserving way, these technologies could enable new insights which \emph{`you might not have been able to gain before because of the sensitivities around the data that you are using'}. P4[P] highlighted a range of \emph{`missed opportunities'} for privacy-preserving computation \emph{`for a good purpose'}. These included cases such as the Boston Women's Workforce Council who \emph{`used secure multiparty computation to confidentially analyze gender wage gaps without ... disclosing who the salary belonged to'}.\footnote{https://thebwwc.org/mpc} P5[P] noted the opportunities for government national security services to use HE techniques like private set intersection to identify suspects without combining certain databases in the clear, something that might not otherwise be undertaken due to the \emph{`intrusiveness'} of sharing data of large numbers of innocent citizens between departments.\footnote{A use case discussed in \cite{de2010practical}.}

% Third, even focusing on the interests of individuals, privacy was often portrayed as just one value among others that needed to be balanced, and data collection could also be of benefit to the individual. As P9(D) explained the mission of her design agency was to work with `public and private organizations to help them improve the way that they using collect data for the benefit of their customers'.

% `So whether that's better protecting people's privacy? And being more transparent about the way that they're using data. And so we've been exploring. How you could use and apply privacy preserving techniques in different contexts to improve people's privacy? Gain new insights, I suppose from from the data that you're collecting, which you might not have been able to gain before because of the sensitivities around the data tht you are using'

While individual privacy was cited by all participants as an important motivator, it was often an indirect motivator, and in some cases perhaps insufficient on its own (e.g. without being coupled with new opportunities to extract value from data). Other participants articulated motivations for pursuing privacy-preserving computation which had nothing to do with individual privacy as such. For example, for some researchers (e.g. P1, P2), it was basic intellectual curiosity (\emph{`somebody thinks of something that ... looks interesting to them'} (P1)). Other cases included where competing businesses would have a mutual interest in the output of some computation on their respective data, but would not otherwise share that data out of \emph{`fear of losing a competitive edge'} (P6[I]). Intellectual property protection was also frequently cited as a key motivation for many business applications.%; this included both privacy of the data being computed on, as well as privacy of the function undertaking the computation (e.g. P6[I] discussed how even if security architecture \emph{`goal is to minimize the risk that an intellectual property and asset is compromised.} P6[I]'. \reuben{find quote for this!\nitin{One option : \textit{The newer technologies, such as homomorphic encryption, multiparty computation, do product data in use, which means that even if to deploy the system you do need the system security architecture built around the system. If that fails and it can fail because its goal is to minimize the risk that an intellectual property and asset is compromised.} P6[I] }}

% The researchers we interviewed were motivated to work on problems of privacy-preserving computation out of inherent intellectual curiosity, and the potential for useful applications to privacy was almost an afterthought:

% \begin{quote}
% `The way technologies are born ... is ... not necessarily directly because of needs. Usually it's just ... somebody thinks of something that ... looks interesting to them and find a way to do it. At least in cryptography ... Yeah it can be used for this and that ... it's a matter of luck and chance' (P1[R])
% \end{quote}

% as a technical replacement for paper protection.

Privacy-preserving computation techniques were also seen by some as offering the possibility to navigate regulatory obligations and trade-offs in different ways. First, they have the potential to fulfill obligations to protect data in new, more `technological' ways, offering \emph{`technological safeguards that can't be easilly overridden'}, the kind of protection that \emph{`paper safeguards, like contractual guarantees and policies, just can't provide'} (P4[P]). They were seen as especially promising in cases where different regulatory obligations might appear to be in conflict, as P4[P] explained:

\begin{quote}
    `Anti money laundering regulations are very data maximalist; they want you to collect more data [to prevent] financial crimes. But in the meantime the GDPR is quite the opposite; it wants you to minimize data, ... and this really conflicts with the regime of AML. I think that PETs could actually cut through these legal conflicts and really provide a practical solution ... it's not actually transferring PII% [Personally Identifiable Information]
    , but it still allows for banks to prepare for AML protocols'
\end{quote}

Similarly, for P5[P], privacy-preserving methods had the capacity to change what is possible without sharing data and thereby shift the scales in legal balancing tests \cite{brown2009terrorism} that might otherwise make certain data analysis unlawful:

\begin{quote}
    ` UK law ... sets out a test for those of us in national security which is \emph{necessity} and \emph{proportionality}. So if you can shift the proportionality, then you're in a better position so you can avoid intruding, you can avoid privacy risk'.
\end{quote}

In these ways, such techniques were envisioned by P4[P] and P5[P] as enabling organisations in the public and private sector to break free of what P4[P] called \emph{`legal gridlocks'} that currently are (or are perceived to) exist around data use and enable new kinds of analysis.\footnote{While a legal analysis of potential conflicts between these two areas of law is beyond the scope of this paper, we note that data protection and financial services regulators have, at least in the UK, affirmed their compatibility in general terms: see \url{https://www.fca.org.uk/news/statements/fca-and-ico-publish-joint-update-gdpr}.}
% `both organizations and individual can gain from them'

% `the first thing that technologists comes with is that, well, you are giving up a little bit of performance. And does performance means usability?
% [People become uncomfortable when you explain that privacy comes at the cost of usability/ performance]'

% `There is also, I would say an important aspect of. Credibility of the field and of homomorphic encryption. Because I I've notice that some organizations are starting to appear that claim they can do homomorphic encryption. They do something else. And and after and some. Oh basically, let's say organizations or customers who are interested in this technology may interact with those organizations or just kind. Of course, starting to sense that there is some hype about homomorphic encryption, and they're trying to use more marketing techniques to get in there and without having the underlying technology or. Uh under basically available to them. I feel like there is there might be some risks that.'
% comparison to blockchain

% E.g. whose privacy, privacy as keeping secrets vs control

% Misconceptions? E.g. output privacy, key management

% Reconfigure tradeoffs

% PETs are for everyone

\subsubsection{Explanation}

Our participants discussed various facets relating to \emph{explaining} privacy-preserving computation, including \emph{how} they go about explaining it to different audiences (and in some cases, why they don't even try).

The researchers described a variety of contexts in which they had had to explain underlying techniques and their strategies for doing so. For a general audience, P2's strategy was to explain simplified versions of protocols, such as simulating a secure multi-party computation for dating using playing cards (see \cite{marcedone2015secure}). While these were \emph{`fun to explain'}, P2[R] was unsure about the effectiveness of such explanations:

\begin{quote}
    `Then in the future, [the audience] will be like: ''Oh yeah, multi-party computation, the thing with the cards.'' That doesn't mean that my explanation was effective... My feeling is that people tend to end up amused and satisfied.' (P2[R])
\end{quote}

Such explanations were offered as a starting point to encourage people \emph{`who are attracted by that kind of magic'} and would \emph{`go into Wikipedia immediately after'} (P2[R]). However, P3[I] felt that there was a lack of accessible educational material: \emph{`there is certainly not enough material and the classical crypto papers are essentially useless for someone who is not an encryption expert'}; they suggested that explanations of core concepts might be more effective if tackled as part of a standardisation process and included within libraries.

Several participants also cautioned that the kind of explanations offered (if any) need to be tailored to the audience. On the one hand, explanations could be too technical: \emph{`If you start with equations ... you lose 99 percent of the audience right away'} (P7[I]). On the other hand, short intuitive explanations might be too simplistic for informing executive decisions:

\begin{quote}
    `So one thing is getting people interested, and the other one is informing, like, executive decisions. I don't think they should be informed ... by two minute stories... I don't think decisions about encryption are made based on an intuitive understanding of crypto'
\end{quote}

For P9[D], designers have a role to play in explaining privacy-preserving techniques through prototyping their use in specific contexts. This included explanations to end users, but also \emph{`a different language to explain it to those designers as well'}.  Previously, their design agency hadn't \emph{`seen much demand for them on the industry side'}; however, that changed after publishing a blog post explaining visually how differential privacy could work in the context of a project on identifying inequalities in urban mobility:

% \begin{quote}
%     Because we focused on that area, it just seemed to resonate with a lot more people than if you just posted a sort of general: `This is randomised response'. Then you have to make the jumps yourself as to how it applies to you. Showing it in context of our real problem and industry as well, you know exactly who you're talking to.
% \end{quote}

\begin{quote}
`Each step of the randomized response process ... we had an image to go with it, so that you could see ... the noise that you are adding to data. %Then when you deployed the technique, what were the privacy preserved results and how do they compare to the raw results. 
Visually seeing it was really helpful for me as a designer and then tying it to sort of real life stories so that I could see how you wouldn't be able to re-identify someone. Imagining what that makes possible forces you to think about the qualities of that technique, what it now enables you to do'
\end{quote}

% `designers and that could help communicate those examples to people, but they I think we also need a different language to explain it to those designers as well so'

When it came to explaining these systems to end-users, however, some participants questioned whether this was a worthwhile goal. P9[D] couldn't imagine \emph{`many scenarios where it's necessary to explain what privacy preserving techniques are being used to an end user who is trying to do something with their phone'}. P7[I] asked himself whether end-users understood these techniques, and answered: \emph{`Well in general, not. Is it a problem? I'm not sure it's a problem'}. In such cases, it was seen as sufficient that end-users \emph{`trust the provider of the solution that they do a good job'}(P7[I]).

\subsubsection{Governance and Accountability}

A final theme was around the challenges of \emph{governance} and \emph{accountability} of privacy-preserving computation. These topics often followed organically from discussions of explanation; attempts to explain these systems were often made in the course of trying to \emph{justify} their use to affected stakeholders, and justification is a key element of accountability \cite{bovens2014public}). But even if explanations don't lead to real understanding on an individual level, it might still be possible to justify them to the public. P5[P] put it this way:

\begin{quote}
`These technologies are extremely difficult to understand... Do they meaningfully address genuine privacy issues? Yes they do. Do they address public concern? That's not to do with the technologies \emph{per se}, [but] how the technologies are explicated and made available. If you told the public: "As a result of using these technologies, we are able to limit the amount of your personal information that's shared, and are still able to offer you valuable services", they would be enthusiasts.'
\end{quote}

Other participants expressed scepticism that the public would take such guarantees at face value. In the context of proposals for privacy-preserving facial recognition in border control, P6 asked: 
\begin{quote}
`if someone publicises this new system ... just by saying: "and by the way the privacy of the information is very well handled because we use the state of the art cryptography", what does that mean to a citizen?'
\end{quote}

Both P6 and P7 suggested that certifications and trust marks applied to services which use these techniques could enable individuals to seek out more trustworthy systems. However, expecting individuals to exercise meaningfully informed choices in relation to different services involving privacy-preserving computation was seen by some as adding to the burden of responsibility unhelpfully placed on individuals. P9 reflected on how \emph{`constantly making decisions about data in the technology that we use is just not sustainable'}; instead, they suggested that \emph{`collective consent models and other governance mechanisms ... that can make decisions on behalf of people'} might be a better approach. Similarly, P2[R] felt decisions about the technical details of the adoption of these technologies ought to be made by \emph{`using experts or authorities'} who can act as \emph{`proxies ... [who] understand their communities'}.

While most of our participants pointed to the positive potential of privacy-preserving computation techniques, a few were also concerned about the power imbalances they might reinforce. When the stakeholders are individuals, they are \emph{`by definition, the weaker party'}, and \emph{`lack the resources ... to induce changes; every time we talk about privacy there is some asymmetry that is implied by it.'} (P1[R]). For P9[D], it is important to recognise the limitations of PPCs as they are just:

\begin{quote}
`a technical solution to protecting people's privacy ... you have to think about the wider system that they sit within and what other kind of power dynamics are in that system.' (P9[D])    
\end{quote}

% \begin{quote}
% `When you talk about HE, the stakeholders are individuals which ... almost by definition, are the weaker party. In this exchange they don't have the resources to pull it together and and and induce changes... Every time we talk about privacy there is some asymmetry that is implied by it.' (P1[R])    
% \end{quote}

% Demystifying / Building trust

% ` there is a business development aspect as well of teaching. Basically, uh, what homomorphic encryption is making sure that potential customers trust the technology and a lot of I would say business and marketing work still needs to be done so that this type of Crypto Technology is perceived similarly. To like a conventional number theory, you know algorithms such I mean like RSA, like more discrete log based algorithms. So I think there is still some substantial work to be done in the, unlike in the business and marketing side.'

\section{Discussion}
The findings from our interviews raised several important implications for the design and governance of privacy-preserving computation. They reveal how these techniques are being not only technically but also socio-technically constructed and constituted by a variety of actors, each pursuing overlapping and sometimes diverging agendas. Clearly, privacy-preserving computation techniques entail a variety of human-centric challenges which HCI research could seek to address. These challenges are multifaceted and will require diverse approaches; something that HCI as a methodologically diverse field is well-positioned to reflect. Furthermore, these challenges are inter-related: for instance, the way in with these technologies are translated from theory to practice may well affect how they can be explained and held accountable; while closer inspection of how `privacy' and other motivations are unpacked might reconfigure what kinds of interdisciplinary collaborations are required in a particular context. Our aim in this section is to reflect on these, to understand both the design problems facing these techniques, and the challenges they raise in relation to the interests of a variety of users and wider society. This discussion is not intended as direct `implications for design'; rather, we hope to draw attention to issues which require further research, as well as interdisciplinary discussion.

% reflect on the findings, orient reader towards whats coming
% While the end user was largely absent from discussions, HCI as a human centred field still has much to contribute to understanding and developing computational PETs in future. In this discussion we aim to outline some of the challenges and opportunities.
% carving the problem space into different areas: there are clear usability challenges facing developers. design of libraries. study how developers learn about and apply their knowledge 
% beset by hype, esoteric and magical connotations, ,

\subsection{Whither the end user?}

% user acceptance has been raised elsewhere:
% e.g. privacy assistant for IoT paper "To what extent does anonymization, aggregation or differential privacy help mitigate people’s reservations about some data collection practices and reduce the chance they opt out?"
While our experts generally acknowledged the individuals whose personal data is being privately computed on as an important stakeholder group, few seemed to prioritise seeking their understanding and acceptance. This is in contrast to the small number of existing HCI studies that investigate `user acceptability' of particular privacy-preserving computation techniques such as differential privacy \cite{bullek2017towards,xiong2020towards} and MPC \cite{qin2019usability}. User acceptability could and should be further examined in particular contexts; for instance, Colnago et al. suggest further work is needed to explore whether such techniques embedded in Internet-of-Things privacy assistants might `help mitigate people's reservations about data collection practices and reduce the chance they opt out' \cite{colnago2020informing}. There is clearly great scope for important research within this paradigm of user acceptability. % However, reflecting on our findings, we suggest that the relative lack of studies on user acceptability in privacy-preserving computation more generally (including SMPC and HE) may not be just because these technologies are relatively new, but rather because they are particularly challenging to study from an end user perspective.

However, our experts spoke about privacy-preserving computation technologies more as tools enabling \emph{organisations} to achieve a variety of goals (including managing privacy risks, but also protecting corporate assets and secrets), rather than as a means of directly serving users' interests. While user acceptance was not entirely disregarded, it did not appear to be a primary concern; even P9, a designer well-versed in user-centred design, doubted that people could or should be expected to understand and make decisions about privacy-preserving computation. Privacy may be important, and these techniques may have the potential to meaningfully embed it, but whether or not individuals understand and accept them seemed to be almost a secondary issue. In many of the use cases they mentioned, individuals whose data is being computed may not have any direct interaction with the system, nor any choice about whether to use it. In expressing such doubts, our interviewees might appear to be denying a sacrosanct tenet of HCI as a \emph{human}-centred discipline. However, rather than denying the importance of user acceptance, we believe that these doubts should in fact point us towards alternative human-centric approaches to the development of privacy-preserving computation, in addition to solely looking at end users as data subjects.

% Speaks to a disjunct between who the technology serves, and whose privacy it aims to protect. Typically, these are one and the same; the user. but here, the tech serves multiple constituents. privacy means several things to each of them, which are often but not always overlapping. the individuals whose privacy is protected are just one such constituent, and for some of our interviewees these individuals should not be expected to concern themselves with how privacy-preserving computation technologies work to protect their interests.

% Those with the most direct interest in and responsibility for their functioning may instead be engineers, developers, designers and risk managers.

First, our findings point towards studying the needs of different kinds of end users; specifically, those developers and designers who attempt to apply foundational privacy-preserving computation techniques in real-world applications. This echoes recent calls to acknowledge that `developers are users too', as Green and Smith argue in relation to crypto and security libraries \cite{green2015developers}. Similarly, P9[D] pointed to the relative lack of awareness and understanding of these techniques among designers. As with the application of other complex methods in computer science, such as machine learning, it may be difficult for designers to use privacy-preserving computation techniques in design practice due to unfamiliarity with how they work and awareness of what they can achieve\cite{dove2017ux}. P9[D] made the case for technical specialists and designers to work together to translate these technologies into `design material' which design practitioners can use to explore real use cases.  % The challenge of working with new or less well understood materials is a recurring theme in some UX research. may not be clear whether PETs are a design material as such.

In addition to understanding developers, designers, and others as \emph{users} of privacy-preserving computation techniques, studying them also allows us to explore how a human value like privacy shapes the construction of complex computational systems. This perspective accords with `third wave' approaches to HCI which orient attention towards the ethical obligations and values of designers \cite{fallman2011new}, and incorporate different disciplinary perspectives which examine how social and political dimensions are embedded and reflected in systems \cite{bardzell2015humanistic}. As such, rather than just considering whether end users or laypeople understand, trust, and accept privacy-preserving computation technologies, we might also benefit from considering the perspectives of the various people involved in constructing their technical, commercial and regulatory foundations. Assessing whether an innovative technology will be acceptable to users through lab and field studies may be valuable, but such approaches often neglect the ways in which such technologies are interpreted, shaped, and mutually constructed over time through their designers, users, and broader political, economic and regulatory forces\cite{mackenzie1999social,jasanoff2009containing}. As a result, it is equally important to consider the plurality of different actors and broader contexts through which values like privacy will be understood, traded-off, and embedded in these systems (or not).

% comment on regulatory implications

% one implication of our findings might be that understanding how PETs affect user's privacy perceptions may not be the most important question. Unlike other privacy tech, it's not clear whether people can or should even begin to understand.

% including researchers, engineers, lawyers, product managers, designers, and government technology advisors, to understand their perceptions of the associated barriers and opportunities, the values and 

% HCI not solely concerned with the 'user', and 'the interface', but considering non-users, and users who are not 'at the interface'. Broader conception of HCI as reflecting human values in computer systems more broadly.

\subsection{The Limits of Abstraction}
If, as suggested above, we are to consider the needs of developers and designers as users of underlying privacy-preserving computation techniques, then how might those needs be met? Many of the interviewees identified the need to create building blocks for privacy-preserving computation. In an ideal world, these building blocks would allow developers to \emph{abstract away} the technical details and apply them to applications in different contexts. Creating such abstractions is fundamental to progress in computer science and programming; in Edsger Dijkstra's words, it is `our only mental aid ... to organize and master complexity' \cite{dijkstra1982selected}.  However, many of the experts expressed uncertainties about the form such abstractions should take and the extent to which they could reasonably be made in the domains of privacy-preserving computation. Especially with homomorphic encryption, abstracting away the details of implementation could mean losing the ability to optimise performance through engineering `tricks' (P1[R]).

Attempts to create tools for developers to enable them to integrate privacy-preserving computation techniques into their products may therefore need to grapple with this need to balance abstraction and engagement with the implementation details. Specific applications will always require some `intimacy with the details' \cite{steimann2018fatal} that might otherwise be abstracted away. Some of our interviewees argued that the necessary education required for developers could potentially be integrated into standardised APIs. This suggests that broader adoption of privacy-preserving computation may benefit from work in HCI which considers APIs and libraries as `first class design objects' \cite{myers2016improving, zibran2008makes}, with the goal of `driving adoption of software components' \cite{mechtley2020api}. This could involve (re)designing them around the typical ways programmers learn, e.g. on-the-fly, via information foraging, and trial and error \cite{lawrance2010programmers,kelleher2019towards}.

However, the nature and extent to which developers need to become intimate with the details, and how they might do so, will clearly depend on the particular technique in question. For instance, a DP library might implement a variety of noise sampling and injection techniques, but this is relatively simple compared to the much more complex mathematics and reasoning involved in deciding on and managing an appropriate privacy budget, which requires case-by-case human consideration. For SMPC, libraries might take care of some of the networking details, but leave difficult decisions regarding the protocol up to the developer. The nature and value of these standardised building blocks will therefore vary greatly between approaches.

Ultimately, the design and adoption of these privacy-preserving computation building blocks may need to reckon with the messy realities of underlying enterprise IT infrastructure, agile and iterative approaches to software development \cite{gurses2017privacy}, and service-oriented architectures \cite{kostova2020privacy}. Given these practical considerations, the full complexity of these technologies might instead need to be mediated via a two-step process: general-purpose libraries which expose all of the complexity of a domain (e.g. homomorphic encryption) that enable specialist privacy engineers to create particular privacy-preserving computation components for common operations or use cases (e.g. private set intersection for contact discovery); those components could then be adapted and deployed with minimal configuration by non-specialist developers as microservices.

\subsection{Privacy-Enhanced Technocracy}

% Will PETs usher in a kind of privacy enhancing technocracy?
The way privacy-enhancing technologies are sometimes described can make them seem esoteric, exotic, and mysterious. For instance, in industry press they have been described as `black magic' \footnote{ https://dualitytech.com/tag/homomorphic-encryption/} and a `holy grail'\footnote{https://www.fastcompany.com/90314942/duality-homomorphic-encryption}. Such language suggests that their development is entirely in the hands of a small and specialised cabal of cryptographers and engineers, much like the early programmers who regarded themselves as `high priests' of assembly code.\footnote{In the words of Rear Admiral Grace Hopper \cite{williams2012grace}} It is possible to imagine how in these respects, they might end up sharing the same `rampant hyperbole and political envisioning' \cite{elsden2018making} of a higher-profile cryptographic technology --- blockchain.

While our interviewees avoided such language, and even criticised the perceived `hype' around PETs, they did reflect the highly specialised knowledge required to make use of the underlying mathematics, and drew parallels with magic. P6[I] described feeling like `Gandalf the wizard' upon telling people that computation on encrypted data was possible,  while P1[R] described the need for engineering `tricks' to optimise performance within reasonable levels. From this perspective, the technical work of applying privacy-preserving computation seems more like craft than science, which the guild of cryptographers and engineers are uniquely capable of performing \cite{prak2006craft}.

However, the mystery of their inner workings could easily serve as an excuse for not making these systems accountable to affected stakeholders. When reflecting on the challenges laypeople face in trying to understand PPCTs on any meaningful level, both P5[P] and P7[I] expressed some doubts about the possibility that individuals could ever be expected to really understand \emph{how} they work. However, without some form of explanation, and absent any other mechanism for meaningfully communicating their risks and opportunities, there is a risk that privacy-preserving computation becomes not just a \emph{technical} but a \emph{technocratic} solution imposed on populations without popular consent by grey eminences operating behind the scenes.

However, several of the experts did acknowledge the need for mechanisms of accountability and governance to developed as these technologies are rolled out. P6[I] and P7[I] suggested this could involve certification schemes. Similarly, while P1[R] and P9[D] were doubtful about individuals being able to meaningfully consent to these technologies, they proposed alternative forms of collective governance, where the interests of affected individuals could be represented by relevant representatives and experts who can make informed choices and demands on their behalf. These and other democratic mechanisms will need to be explored in order to counter a privacy-enhanced technocracy, and methods from HCI --- such as participatory design~\cite{ehn1988work}, futures workshops \cite{jungk1987future}, and other governance approaches --- may have much to offer.

\subsection{Secrets, Assets, Human Rights: Unpacking `Privacy'}
Our findings attest to the many varied interpretations and uses of the term privacy. As previous work has explored, and as discussed above, the notion of privacy in Privacy-Enhancing Technologies is often a narrow interpretation of what is a multi-faceted and contested concept \cite{tavani2001privacy,gurses2010pets,stalder2002failure,phillips2004privacy}. This is certainly the case for the subset of privacy-preserving computation PETs studied here. They turn privacy into something mathematically formalisable, e.g. in terms of entropy in cryptographic approaches, or indistinguishability in statistical approaches, which can all be understood as variations of `confidentiality', a pillar of the security triad \cite{dhillon2001current}. This means that other ways of understanding privacy may be de-emphasised and de-prioritised.

There are continuities here with earlier PETs, such as de-identification techniques based on hashing personal data. Phillips argues that that these techniques embody privacy as protection `from unwanted intrusion'\cite{phillips2004privacy}. However, they leave in place the ability of powerful observers to produce `panoptic' knowledge which can be used to sort and discipline populations \cite{gandy1993panoptic, cohen2012configuring}. Similarly, if we understand privacy as confidentiality, this can be engineered through architectures of data minimisation \cite{spiekermann2008engineering}; but this can lead to design choices which preclude alternative understandings of privacy (e.g. privacy-as-control), and hinder the exercise of related rights afforded by data protection law \cite{veale2018data}. In our expert's discussions, these alternative understandings of privacy were conspicuous by their absence.

%  An example of the latter would be approaches which prevent multiple data points pertaining to an individual being easily linked between databases, but as a result limit that individual's ability to access to their data as per Article 15 of the GDPR \cite{veale2018data}.

Our findings also demonstrate that even while discourse around privacy-preserving computation restricts certain interpretations of privacy, it also stretches the meaning of privacy to incorporate unorthodox meanings, such as competitive secrecy, corporate asset protection, and government security. These are clearly significant and important use cases for the technology, but they arguably bear only a family resemblance to privacy as it relates to individuals and society. Indeed, intellectual traditions which value privacy as an individual right and public good have often been associated with opposition to corporate and government secrecy; according to them, privacy should be reserved for the weak, while transparency should be an obligation required of the strong \cite{de2006privacy}. In referring to all of these things as `privacy', privacy-preserving computation technologies may elide significant political tensions between them. This is not to deny that they may have a powerful role to play in supporting privacy as an individual right and as a public good \cite{kwecka2014spartacus,fairfield2015privacy}; but this confluence of quite different values under one banner complicates the narrative around whose interests they serve.

As well as tending to address narrow and perhaps unorthodox conceptualisations of privacy, it is important to recognise that these technologies do not protect other important values and interests. If our aim is to build and shape systems encompassing multiple social goals, where privacy is just one such goal, then privacy-preserving computation techniques have to be considered in relation to the whole system and the social context. The danger is that the societal problems of data processing technologies --- such as the ways they create distinctions and hierarchies that reinforce power, shape politics, or facilitate abuse --- are sidelined, redefined, or collapsed under the banner of `privacy', so that privacy-preserving computation techniques can be positioned as \emph{the} solution (what Pinch and Bijker term `closure by problem redefinition' \cite{pinch1984social}). This danger was alluded to in P6[I]'s example of privacy-preserving border control (where people are still ultimately at the mercy of a powerful state), and in P9[D]'s concern about considering the wider power dynamics in the context of deployment.

\section{Conclusions}

Like other technologies which have been touted as potentially revolutionary in recent years, the concrete impact of these privacy-preserving computation techniques remains to be seen. New technologies often emerge in unexpected ways, at unpredictable times from niches of computer science: hypertext, Merkle trees, and neural networks were once confined to their respective research sub-fields before they became known more widely as the world wide web, blockchain, and `AI' (in its latest guise of deep learning). Prior to their take-up in wider society, these specialised areas of research were conceived as laying the groundwork for purely technical pieces of invisible infrastructure, whose implications for human-computer interaction were remote and unclear.

However, we believe it is worth HCI researchers studying such technologies prior to their widespread adoption. Whatever technical and institutional forms they take, the journey of privacy-preserving computation techniques from the annals of cryptography into production code will be shaped in substantial part by the approach they take to a variety of human and societal challenges. Indeed, these challenges directly implicate some fundamental concerns of HCI, including: multifaceted (re)conceptualisations of the notion of `the user'; helping people navigate and manage computational complexity and its consequences; exploring how values like privacy can be reflected in the systems we build; and examining how different political agendas, economic rationale, and user groups shape and are shaped by those systems. These concerns all cohere and overlap in the emerging space of privacy-preserving computation.

This paper has aimed to  provide a preliminary and partial outline of those challenges, laying some of the groundwork for substantial further exploratory and in-depth work to be done. In addition to several recent studies which focused on people's understanding of these techniques and their willingness to disclose personal data in the presence of them, we have outlined a broader set of research questions prompted for HCI by PPCs. These include understanding specific application contexts; usability of PPC libraries and tools from a non-specialist developer's perspective; and understanding the explanation and governance challenges associated with these techniques.

% We hope this lays some groundwork for a broader research agenda in this area, including examples of potential specific studies to investigate the barriers to adoption identified by our participants, such as developer usability, explainability, and governance

% make this point in relation to other areas of advanced CS: As with the application of other complex methods in computer science, such as machine learning, it may be difficult for designers to use privacy-preserving computation techniques in design practice due to unfamiliarity with how they work and awareness of what they can achieve
% a role for HCI as a mediator between complex CS and the messy realities of the world

%  `While direct manipulation continues to be a strong ideal for much work in HCI to this day, it is questionable whether it can continue to function as a unifying, agreed-upon ‘good’, especially when taking into account the current broadening of HCI’s scope, the new areas where HCI thinking is applied, and the multifaceted notion of the user in third wave HCI' \cite{fallman2011new}. this is also true for HCI and privacy, and computational PETs are a good illustration of why.

\begin{acks}
% This work was funded by:
% PETRAS Grant (?)
% Callsign Inc. 
This work was funded by EPSRC grant EP/S035362/1 and Callsign Inc.
\end{acks}

% \nitin{Double check protocols in the Appendix}

\bibliographystyle{ACM-Reference-Format}
\bibliography{refs}

% \appendix
% \input{App_Protocol_Experts}
% \input{App_Protocol_policy_design}
\end{document}